\documentclass[conference]{IEEEtran}
\ifCLASSINFOpdf
\else
\fi
\usepackage{amssymb}
\usepackage{amsthm}
\usepackage{tabulary}
\usepackage{steinmetz} 
\usepackage{multirow}
\usepackage{textcomp} 
\usepackage{gensymb}
\usepackage{color}
\usepackage[font=scriptsize]{subcaption}
\usepackage{floatrow}
\usepackage{graphicx}
\usepackage{epstopdf}
\usepackage{rotating}
\usepackage{bigstrut}
\usepackage[disable]{todonotes} 
\newcolumntype{C}[1]{>{\centering\let\newline\\\arraybackslash\hspace{0pt}}m{#1}}
\graphicspath{/pics}

\usepackage{graphicx}
\usepackage{amsmath,amsfonts,amssymb}
\usepackage[printonlyused, nolist]{acronym}
\usepackage{listings}      
\usepackage{xcolor}

\usepackage{tikz}
\usepackage{colortbl}

\definecolor{clight2}{RGB}{89, 181, 175}%
\definecolor{glight2}{RGB}{212, 212, 212}%
\definecolor{Blue}{RGB}{31, 119, 180}%
\definecolor{Green}{RGB}{89, 181, 175}%
\definecolor{Red}{RGB}{37, 25, 68}%
\newcommand\tikznode[3][]%
   {\tikz[remember picture,baseline=(#2.base)]
      \node[minimum size=0pt,inner sep=0pt,#1](#2){#3};%
   }
\tikzset{>=stealth}

\makeatletter
\renewcommand*\env@matrix[1][*\c@MaxMatrixCols c]{%
  \hskip -\arraycolsep
  \let\@ifnextchar\new@ifnextchar
  \array{#1}}
\makeatother

\begin{document}
%
\title{An efficient open-source implementation to compute the Jacobian matrix for the Newton-Raphson power flow algorithm}

\author{\IEEEauthorblockN{Florian Sch\"afer}
\IEEEauthorblockA{University of Kassel\\
Wilhelmsh\"oher Allee 73, 34121 Kassel, Germany\\
Email: florian.schaefer@uni-kassel.de}
\and
\IEEEauthorblockN{Martin Braun}
\IEEEauthorblockA{University of Kassel\\Fraunhofer Institute for Energy Economics\\ and Energy System Technology\\
Koenigstor 59,  34119 Kassel, Germany\\
martin.braun@iee.fraunhofer.de}}


%


\maketitle
\lstset{language=Python}
\renewcommand{\lstlistingname}{Pseudocode}
\begin{abstract}
Power flow calculations for systems with a large number of buses, e.g. grids with multiple voltage levels, or time series based calculations result in a high computational effort. A common power flow solver for the efficient analysis of power systems is the Newton-Raphson algorithm. The main computational effort of this method results from the linearization of the nonlinear power flow problem and solving the resulting linear equation. This paper presents an algorithm for the fast linearization of the power flow problem by creating the Jacobian matrix directly in \ac{CRS} format. The increase in speed is achieved by reducing the number of iterations over the nonzero elements of the sparse Jacobian matrix. This allows to efficiently create the Jacobian matrix without having to approximate the problem. A comparison of the calculation time of three power grids shows that comparable open-source implementations need 3-14x the time to create the Jacobian matrix.
\end{abstract}


%
\IEEEpeerreviewmaketitle

\section{Introduction}
\label{intro}
Power flow studies are the basis for operating, planning and analysing power systems \cite{Grainger.1994}. Algorithms for the computation of power flows are being developed since more than 50 years \cite{Tinney.1967} with the focus on improving the convergence behavior \cite{Rao.2015} or reducing the computational effort \cite{Araujo.2006}\cite{Semlyen.2001}.  The efficiency of the solver is crucial if systems with a large number of nodes need to be analyzed (e.g, transmission and distributions grids including several voltage levels). Also if numerous power flow calculations are necessary, the computational time for each calculation needs to be minimized. This is especially relevant in planning studies, where a lot of different network configurations must be tested \cite{Ahmadi.2015} \cite{MartinezCesena.2016b}.

A standard approach for solving the power flow problem is the Newton-Raphson method, which is proven to be robust and efficient at the same time. This method finds a solution of the nonlinear power flow problem in an iterative manner. During each iteration, the power flow problem is linearized, and the resulting system of linear equations is solved. For a system with $n$ buses, the typical complexity of the algorithm is O($n^2$) \cite{Tinney.1967}. The main computational effort results from two aspects: First, the linearized problem - defined by the Jacobian matrix and power mismatches - needs to be formulated. Second, the linear problem has to be solved. Fast solvers for systems of linear equations are widely available \cite{Li.1999} \cite{Davis.2010}. However, the time needed for the formulation of the linear problem (i.e., the construction of the Jacobian matrix), and its solution strongly depends on the implemented algorithm and the programming environment. Especially the creation of large matrices and necessary mathematical operations on these are critical in terms of speed. 

\subsubsection{Available and comparable open-source Newton-Raphson implementations}
Common open source implementations, such as \textsc{pypower} \cite{Lincoln.2015} and \textsc{matpower} \cite{Zimmerman.2011}, are based on generic functions to calculate the Jacobian matrix. \textsc{matpower} relies on the internally available generic functions for stacking an creating large compressed sparse matrices delivered with \textsc{matlab}. Similarly, \textsc{pypower} is based on \textsc{numpy} \cite{Jones.2001} and comparable functions for stacking and creating large compressed sparse matrices. These generic functions are easy to use for developers, but have the disadvantage of iterating over the sparse matrices more often than actually necessary to compute the power derivatives. For large matrices or in cases where these matrices have to be computed very often, tailored algorithms are able to significantly reduce the calculation time.

\subsubsection{Aim of this paper}
This paper presents a fast method to calculate the Jacobian matrix based on the \ac{CRS} storage format. By exploiting the characteristics of the sparse admittance and Jacobian matrix, the number of iterations over the nonzero elements of these matrices is minimized and the computational effort is reduced. The algorithm is implemented in the open source power system analyzing tool pandapower \cite{Thurner.2017} and compatible to \textsc{pypower} \cite{Lincoln.2015} as well as \textsc{matpower} \cite{Zimmerman.2011}. Pandapower is a Python based module that combines the data analysis library \textsc{pandas} \cite{McKinney.2010} and power flow solvers to create an easy to use network calculation framework. The implemented Newton-Raphson power flow solver is originally based on \textsc{pypower} \cite{Lincoln.2015}, but includes several performance and convenience improvements. One of these improvements is the developed algorithm to create the Jacobian matrix, which is presented in this paper.

The remainder of this paper is structured as follows: Sections \ref{sec:NR} and \ref{sec:CRS} describe the basics to understand the Newton-Raphson method and the applied sparse matrix storage format of the algorithm. In section \ref{sec:J} the algorithm is explained in detail. Speed comparisons are shown in section \ref{sec:comparison}. In the last section a conclusion and outlook is given.
%
\section{Newton-Raphson power flow method}
\label{sec:NR}
The Newton-Raphson power flow algorithm is an iterative method, based on the linearization of the power flow problem. Starting from an initial solution, the calculated injected power at every bus in a system is being updated in every step. For this, the linear problem ${J} x = [\Delta P, \Delta Q]$ (eq. \eqref{eq:linProblem}) is formulated and solved in every Newton-Raphson iteration \cite{Tinney.1967}. 

Starting from the initial voltage estimate, the Newton-Raphson method updates this estimation iteratively until the method is converged or a maximum number of iterations is reached. Convergence is achieved if the mismatch between the scheduled power injections $S_{i,s} = P_{i,s} + jQ_{i,s}$ and the calculated injections $S_i = P_i + jQ_i$ at every non-slack bus are smaller then a defined tolerance $\epsilon$. The injections $S_i$ are derived from the current voltage estimate \cite{Grainger.1994}. The difference between the scheduled injection and the calculated estimate is expressed by the mismatch equations \eqref{eq:dP} and \eqref{eq:dQ}:
\begin{equation}
\Delta P_i = P_i - P_{i,s}  
\label{eq:dP}
\end{equation} 
\begin{equation}
\Delta Q_{i} = Q_i - Q_{i,s}
\label{eq:dQ}
\end{equation}
with
\begin{equation}
P_i = \sum_{k=1}^N |V_i||V_k|(G_{ik}\cos\theta_{ik}+B_{ik}\sin \theta_{ik})
\end{equation}
\begin{equation}
Q_i = \sum_{k=1}^N |V_i||V_k|(G_{ik}\sin\theta_{ik}-B_{ik}\cos\theta_{ik})
\end{equation}
$P_i$ and $Q_i$ are the net real and reactive power injections at bus $i$ with the voltage magnitude $|V_i|$. Similarly, $|V_k|$ is the voltage magnitude at bus $k$. Between bus $i$ and $k$ the difference in the voltage angles is $\theta_{ik} = \delta_{i} - \delta_{k}$. Taylor series are used to linearize the power flow problem, which then can be expressed in matrix form as:
\begin{equation}
J \begin{bmatrix} \Delta V_a \\ \Delta V_m\end{bmatrix} = \begin{bmatrix}\Delta P \\ \Delta Q \end{bmatrix}
\label{eq:linProblem}
\end{equation}  
where ${J}$ is called the Jacobian matrix, which contains the partial derivatives of $\Delta P$ and $\Delta Q$ with respect to the voltage angle $V_a$ and magnitude $V_m$ \cite{Grainger.1994}. 
\begin{equation}
{J}=\begin{bmatrix} \dfrac{\partial P}{\partial V_a} & \dfrac{\partial P}{\partial V_m} \\ \dfrac{\partial Q}{\partial V_a}& \dfrac{\partial Q}{\partial V_m}\end{bmatrix} = \begin{bmatrix} J_{11} & J_{12} \\ J_{21} & J_{22} \end{bmatrix}
\label{eq:J}
\end{equation}
For every PQ-bus the partial derivatives $\frac{\partial P}{\partial V_a}$, $\frac{\partial Q}{\partial V_a}$, $\frac{\partial P}{\partial V_m}$ and $\frac{\partial Q}{\partial V_m}$ need to be calculated in each iteration. Similarly, the partial derivatives $\frac{\partial P}{\partial V_a}$ and $\frac{\partial Q}{\partial V_a}$ have to be calculated for every $PV$ bus.

\section{Compressed Row Storage Format}
\label{sec:CRS}
Buses in realistic power systems are connected to only a few other buses. Thus, the admittance and the Jacobian matrix have a high sparsity and are typically stored in sparse matrix storage formats. 

The \ac{CRS} scheme stores subsequent nonzero elements of a matrix ${A} = (a_{ij}) \in \mathbb{R}^{n \times n} $ in contiguous memory locations. Instead of storing every entry of the matrix explicitly, the \ac{CRS} Format is based on three vectors. These vectors store the positions and values of the nonzero elements. The first (data) vector $A_x$ stores only the nonzero values of $A$ as floating point numbers. A value at position $k$ in the array is accessed by an index operator $[$ $]$:
\begin{equation}
a_{ij} = A_x[k]
\end{equation}
The second vector $A_j$ contains the column indices of the entries in $A_x$ as integers:
 \begin{equation}
A_j[k] = j
\end{equation}
The third vector (row pointer) $A_p$ contains also integers, which store the locations in $A_x$ that start a row:
\begin{equation}
A_p[i] \leq k < A_p[i+1]
\end{equation}
By convention,  $A_p(n+1) = \text{nnz} + 1$ is defined, where \ac{nnz} is the number of nonzeros in $A$. The storage savings are greater when the sparsity of $A$ is higher. Instead of storing $n^2$ elements, only $2\text{nnz} + n + 1$ values need to be stored \cite{Barrett.1994b}. 

\section{Algorithm to calculate the Jacobian matrix}
\label{sec:J}
Common open source implementations \cite{Lincoln.2015}\cite{Zimmerman.2011} determine the Jacobian matrix by:
\begin{enumerate}
\item Calculating the partial derivatives (see eq. \eqref{eq:J}) as matrices for every non-slack bus
\item Creating the Jacobian matrix in \ac{CRS} format by selecting values from the derivatives matrices
\end{enumerate}
These steps need to be repeated in every Newton- Raphson iteration and are time consuming, especially for systems with more than a few hundred nodes. A tailored implementation of these steps, which exploits the sparsity of the matrices, is presented in this paper. In the following, $\partial V_m$ is defined as the matrix which contains the partial derivatives of $P$ and $Q$ with respect to $V_m$. Similarly, $\partial V_a$ is defined as the matrix which contains the partial derivatives of $P$ and $Q$ in respect to $V_a$.

\subsection{Calculate derivatives}
The partial derivatives of the voltage magnitudes $\partial V_m$ and the voltage angles $\partial V_a$ are determined by eq. \eqref{eq:dS_dVm} and \eqref{eq:dS_dVa} (see \cite{Tinney.1967} for details). Where $d()$ is defined as an operation, which creates a diagonal matrix from its containing vector: 
\begin{equation}
{\partial V_m} = d(V) \cdot (\overbrace{{Y} \cdot d(V_{norm})}^{\rightarrow \eqref{eq:dVm1} })^* + \overbrace{d(I)^* \cdot d(V_{norm})}^{\rightarrow \eqref{eq:dVm2}}
\label{eq:dS_dVm}
\end{equation}
\begin{equation}
{\partial V_a} = j d(V) \cdot (d(I) - \overbrace{{Y} \cdot d(V)}^{\rightarrow \eqref{eq:dVa1}})^*
\label{eq:dS_dVa}
\end{equation}
with
\begin{equation}
\partial V_a, \partial V_m \in \mathbb{C}^{n \times n}
\end{equation}
\begin{equation}
V = \begin{bmatrix}
v_{1} & \cdots & v_{n} \\
\end{bmatrix} \in
 \mathbb{C}^{1 \times n}
\end{equation}
\begin{equation}
I =
\begin{bmatrix}
y_{11} & \cdots & y_{1n} \\
\vdots  & \ddots & \vdots  \\
y_{n1} & \cdots & y_{nn} \\
\end{bmatrix} 
\begin{bmatrix}
v_{1} \\ \vdots \\ v_{n} 
\end{bmatrix}
=
\begin{bmatrix}
i_{1} \\ \vdots \\ i_{n} 
\end{bmatrix}
= Y \cdot V
\in \mathbb{C}^{n \times 1}
\label{eq:I}
\end{equation}
\begin{equation}
V_{norm} = \frac{V}{V_m} = (v_{norm,i}) \in \mathbb{C}^{1 \times n} 
\end{equation}

Equations \eqref{eq:dS_dVm} and \eqref{eq:dS_dVa} require matrix dot products, additions and conjugations. Generic implementations of these sparse matrix operations have to iterate over the \ac{nnz} elements of the matrices once for each operation. To calculate equations \eqref{eq:dS_dVm}, \eqref{eq:dS_dVa} and \eqref{eq:I}, it is necessary to iterate over the \ac{nnz} elements of $Y$ 6 times for multiplications, twice for additions and 3 times for conjugations. In total $11 \cdot nnz$ iterations are necessary. 

With the proposed algorithm in this paper, it is possible to combine multiple operations in a total of two iterations over the \ac{nnz} elements in $Y$. This can be achieved, since the resulting matrices $\partial V_m$ and $\partial V_a$ have the same dimension as $Y$. Additionally, the computational effort is reduced by only creating the data vectors of $\partial V_m$ and $\partial V_a$, instead of recalculating the row pointers and column indices. This is possible since the \ac{nnz} as well as the column indices and the row pointer are identical for $Y$, $\partial V_m$ and $\partial V_a$. Thus, only the data vectors $\partial V_{m,x}$ and $\partial V_{a,x}$ differ from $Y_x$ and need to be computed to generate the \ac{CRS} representation of the matrices $\partial V_m$ and $\partial V_a$. Pseudocodes \ref{pc:dS1} and \ref{pc:dS2} describe the implementation for the calculation of the data vectors $\partial V_{m,x}$ and $\partial V_{a,x}$  with loops: 
 
\begin{lstlisting}[frame=single, mathescape, caption=Calculation of derivatives 1, label=pc:dS1] 
for i in rows of Y:
  for k in nonzero elements per row:
    I[i] = I[i] + $Y_x$[ik] $\cdot$ V[k]
    $\partial V_{m,x}$[ik] = $Y_x$[ik] $\cdot$ $V_{norm}$[k]
    $\partial V_{a,x}$[ik] = $Y_x$[ik] $\cdot$ V[k]
  end
  temp[i] = I[i$]^*$ $\cdot$ $V_{norm}$[i]
end
\end{lstlisting}
In these loops, equation \eqref{eq:I} is calculated:
\begin{equation}
i_{i} = \sum_{k=1}^n y_{ik} \cdot v_{k}
\end{equation}
as well as the following parts of \eqref{eq:dS_dVm} and \eqref{eq:dS_dVa}:
 \begin{equation}
 \partial V_{m,x,ik} = y_{ik} \cdot v_{norm,k}
 \label{eq:dVm1}
 \end{equation}
 \begin{equation}
 \partial V_{a,x,ik} = y_{ik} \cdot v_{k}
 \label{eq:dVa1}
 \end{equation}
\begin{equation}
temp_{i} = i_{i}^*  \cdot v_{norm,i}
 \label{eq:dVm2}
\end{equation}
The resulting vectors from Pseudocode \ref{pc:dS1} are the input to Pseudocode \ref{pc:dS2}. With these vectors, the data vectors of the derivatives are created:
\begin{lstlisting}[frame=single,mathescape, caption=Calculation of derivatives 2, label=pc:dS2] 
for i in rows of Y:
  for k in nonzero elements per row:
    $\partial V_{m,x}$[ik] = $\partial V_{m,x}$[ik$]^*$ $\cdot$ V[i]

    if diagonal element:
      $\partial V_{m,x}$[ik] = $\partial V_{m,x}$[ik] + temp[i]
      $\partial V_{a,x}$[ik] = I[i] - $\partial V_{a,x}$[ik]
    end
 
    $\partial V_{a,x}$[ik] = $\partial V_{a,x}$[ik$]^*$ $\cdot$ jV[i]
  end
end
\end{lstlisting}
Pseudocode \ref{pc:dS2} is an implementation of the following equations:
\begin{equation}
\partial V_{m,x,ik} = v_i \cdot (\partial V_{m,x,ik} )^* + temp_{i}
\end{equation}
\begin{equation}
\partial V_{a,x,ik} = j v_i \cdot (i_i - (\partial V_{a,x,ik}))^*
\end{equation}

$\partial V_{a,x}$, $\partial V_{m,x}$ together with $Y_p$ and $Y_i$ represent $\partial V_m$ and $\partial V_a$ in \ac{CRS} format.
Although the amount of mathematical operations stays the same as with the generic implementation, the number of iteration steps over the \ac{nnz} elements are reduced from $11 \cdot nnz$ to one iteration in each loop.

\subsection{Creation of Jacobian matrix in \ac{CRS} format}
The Jacobian matrix is filled with parts of the voltage derivatives matrices $\partial V_m$ and $\partial V_a$. For this, the arrays $pv$ and $pq$ are defined, which are masks to index parts of these matrices. These arrays contain the bus indices of the PV- and PQ-buses of a grid with $N_{pv}$ PV- and $N_{pq}$ PQ-buses. $N_{pvpq} = N_{pv} + N_{pq}$ equals the sum of the number of PV and PQ-buses.  Together with an index operator $[$ $]$, the masks $pv$ and $pq$ select the imaginary and real parts of the matrices $\partial V_m$ and $\partial V_a$  to create following matrices:
\begin{equation}
J_{11} = \Re(\partial V_a [pvpq, pvpq]) 
\end{equation}
\begin{equation}
J_{12} = \Re(\partial V_m[pvpq, pq]) 
\end{equation}
\begin{equation}
J_{21} = \Im(\partial V_a[pq, pvpq]) 
\end{equation}
\begin{equation}
J_{22} = \Im(\partial V_m[pq, pq]) 
\end{equation}
and stacking them to obtain the full Jacobian matrix:
\begin{equation}
J = \begin{bmatrix} J_{11} & J_{12} \\ J_{21} & J_{22} \end{bmatrix} \in \mathbb{R} 
	\begin{bmatrix} [c|c]
	N_{pvpq} \times N_{pvpq} & N_{pvpq} \times N_{pq} \\ 
	\hline
	N_{pq} \times N_{pvpq} & N_{pq} \times N_{pq}
	\end{bmatrix} 
\end{equation}
The selection and stacking of the sub-matrices is avoided by directly creating the row pointer $J_p$, the column indices $J_i$ and the data vector $J_x$ of the Jacobian matrix with the following steps:
\begin{enumerate}
\item Count the total \ac{nnz} in $J$
\item Iterate over the rows of $J$, which equal $N_{pvpq}$
\item For the current bus in pvpq iterate over the columns $V_j$ for the row in $V_p$
\item If an entry exists for the current bus in V, write an entry for $J_x$ and $J_j$
\item Count the \ac{nnz} entries in current row and add them to the row pointer $J_p$
\end{enumerate}
Pseudocode \ref{pc:J} exemplary describes the algorithm for creating the entries of $J_{11}$ and $J_{12}$:

\begin{lstlisting}[language=Python, numbers=left, xleftmargin=2em,frame=single,framexleftmargin=1.7em,mathescape, caption= Creation of CRS vectors for $J_{11}$ and $J_{12}$, label=pc:J] 
nnz = 0
for row in 0 to length(pvpq):
  nnz_row = nnz
  bus = pvpq[row]
  for k in $Y_p$[bus] to $Y_p$[bus+1]:
    j = $bus_{indices}$[$Y_i$[k]]
    if pvpq[j] == $Y_i$[k]:
      # entry for $J_{11}$
      $J_x$[nnz] = $\partial V_{a,x}$[k].real
      $J_i$[nnz] = j
      nnz += 1
      if j >=  $N_{pv}$:
        # entry for J12
        $J_x$[nnz] = $\partial V_{m,x}$[k].real
        $J_i$[nnz] = j + $N_{pv}$
        nnz += 1
      endif
    endif
  end
  $J_p$[row+1] = nnz - nnz_row + $J_p$[row]
end     
\end{lstlisting}
Similarly, the entries of $J_{21}$ and $J_{22}$ are written to $J_x$, $J_p$ and $J_i$. This requires an additional iteration over the number of rows ($N_{pq}$)  and the corresponding nonzero elements in $\partial V_{m}$ and $\partial V_{a}$ of these rows. In total it is necessary to iterate over the \ac{nnz} elements of $N_{pv} + 2 \cdot N_{pq}$ rows in $\partial V$.

\section{Comparison of calculation time}
\label{sec:comparison}
Three different \textsc{matpower} power system cases, available from \cite{Zimmerman.2011}, are analysed to benchmark the calculation time. Since the number of buses correlate with the computational effort, the cases with 118, 1354 and 9421 are chosen to show results for different grid sizes. A flat start ($V_m = 1.0~p.u.$ and $V_a = 0^\circ$) is the initial solution for the Newton-Raphson method for each case.
\subsection{Numba \acs{jit}-compiler vs. pure Python}
Even though a specific implementation reduces the computational effort compared to a generic implementation, it is still crucial to have an efficient resource management to achieve a low computational time. The algorithm implemented in pandapower is entirely written in the Python programming language. \textsc{pypower} and \textsc{matpower}, however, use pre-compiled algorithms (\textsc{numpy}, internal \textsc{matlab} functions) for the calculation of the derivatives and stacking of the sub-matrices of the Jacobian matrix. Since the overall computational speed is to be reduced, the native Python code is converted to machine instructions with the \ac{jit}-compiler \textsc{numba} \cite{Lam.2015}. 

\begin{figure}[ht]
  \includegraphics[trim=0 0 0 0,clip, width=1.0\textwidth]{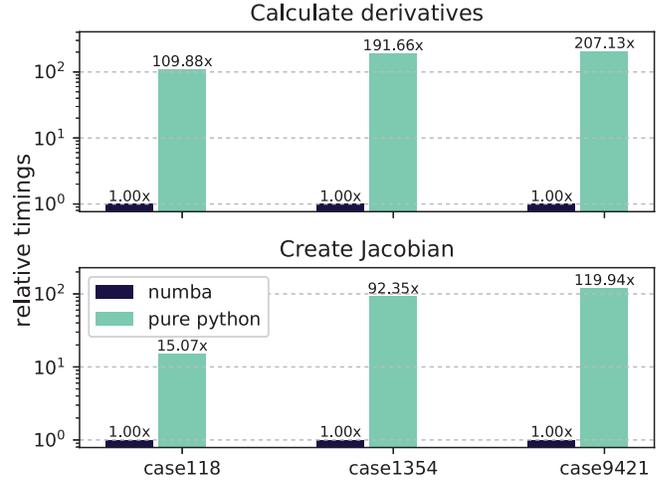}
\caption{Calculate derivatives and creating of Jacobian matrix - Numba vs. pure Python}
\label{fig:numba_vs_native}       
\end{figure}

To outline the speed difference, Fig. \ref{fig:numba_vs_native} shows the computational time of the pure Python implementations related to the computational time of compiled Numba code for the developed algorithm. 

It can be seen, that the higher the quantity of nodes the greater is the relative speed difference. Especially in case9421, a pure Python implementation of the algorithm needs more than 207x the time to calculate the derivatives and 119x the time to create the Jacobian matrix. Because of the significant difference in computational speed, only the Numba compiled versions in combination with pandapower are analysed in the following comparisons.

\subsection{Comparison of the algorithms implemented in pandapower, \textsc{pypower} and \textsc{matpower}}
In this section three different implementations of the Newton-Raphson solver are compared:
\begin{enumerate}
\item The algorithm presented in this paper, implemented in pandapower v1.2.2 in combination with \textsc{numba} v31.0
\item The implementation from \textsc{pypower} v5.0.1 in combination with \textsc{numpy} v1.11
\item The implementation from \textsc{matpower} v6.0b2 in \textsc{matlab} 2015b
\end{enumerate}
Every implementation is available as open source software \cite{Thurner.2017}\cite{Lincoln.2015}\cite{Zimmerman.2011}. In the following figures, the computational time of the corresponding software versions are referred as "pandapower", "pypower" and "matpower". The comparisons show the shortest calculation times of 100 sequential power flow runs of the Newton-Raphson solver implementations. This minimizes the influence of other processes running on the benchmark system and eliminates the initial setup costs for each environment. The conversion overhead of the input data is not compared for the three tools. The calculations were computed on an Intel Core i7-4712MQ CPU @ 2.30GHz with 16GB RAM on Windows 7 64-bit.
 
\begin{figure}[ht]
  \includegraphics[trim=0 0 0 0,clip, width=1.0\textwidth]{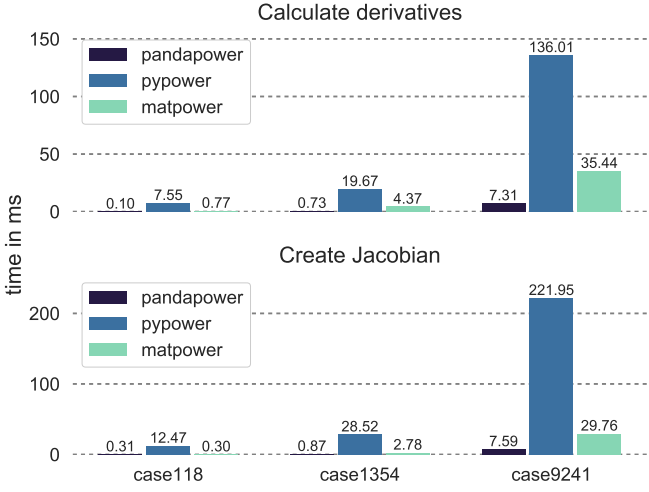}
\caption{Computational time to calculate the derivatives and to create the Jacobian matrix}
\label{fig:numba_vs_numpy_vs_matlab}       
\end{figure}

\paragraph{Calculation of derivatives and creation of the Jacobian matrix}
In Fig. \ref{fig:numba_vs_numpy_vs_matlab} the absolute time in ms of the calculation of the derivatives $\partial V_m$ and $\partial V_a$ (top) and the creation of the Jacobian matrix (bottom) for each implementation are shown.

In relation to pandapower and \textsc{matpower}, the \textsc{pypower} implementation needs longer to calculate the derivatives and to create the Jacobian in every analysed case. The main computational effort results from the creation of the diagonal matrices from the vectors (eq. \eqref{eq:dS_dVm} and \eqref{eq:dS_dVa}) and the mathematical operations on these matrices. Also the stacking of the partial Jacobian matrices is slower compared to the other implementations. In dependency of the number of buses, the calculations in \textsc{pypower} need more than 20x the time in relation to the pandapower implementation. Also the \textsc{matpower} implementation takes up to 4-7x the amount of time compared to the pandapower version, which includes the presented algorithm.

\begin{figure}[ht]
  \includegraphics[trim=0 0 0 20,clip, width=1.0\textwidth]{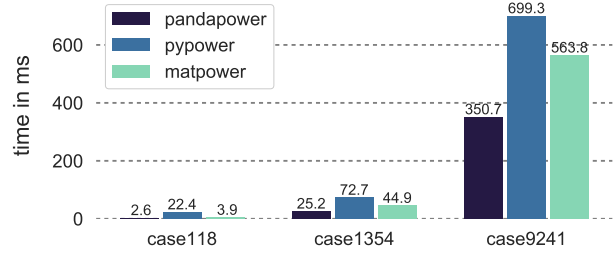}
\caption{Computational time of the Newton-Raphson solver in total}
\label{fig:nr_total_timings}       
\end{figure}

\paragraph{Total computational time of the Newton-Raphson solver}
In Fig. \ref{fig:nr_total_timings} the total computational time of the Newton-Raphson solvers are compared. In case118 the \textsc{pypower} implementation needs 8.6 times as long as the Numba compiled pandapower version to find a solution. Also for the calculation of cases 1345 and 9241, the \textsc{pypower} implementation needs more than twice the time even though the same linear solver is used. \textsc{matpower} is generally faster than the \textsc{pypower} version and slower than pandapower for the analysed cases. For the calculation of the case with the highest quantity of buses (9241), \textsc{matpower} needs 1.6x the time to find a solution in relation to pandapower. 

\begin{figure}[ht]
  \includegraphics[trim=0 0 0 20,clip, width=1.0\textwidth]{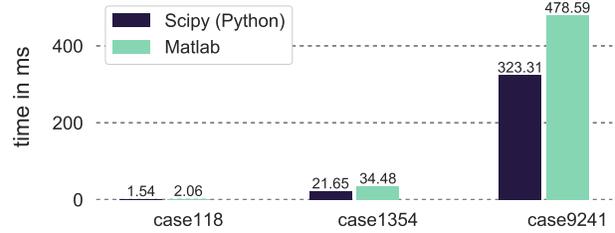}
\caption{Computational time of the linear solver}
\label{fig:solve_linear_system}       
\end{figure}

\paragraph{Linear solver}
The computational time shown in Fig. \ref{fig:nr_total_timings} do not explicitly portray the time needed by the linear solver. It must be noted, that the linear solver implemented in \textsc{matlab} is based on \textsc{umfpack} \cite{Davis.2010}, whereas \textsc{scipy} (Python) uses SuperLU as the default solver \cite{Li.1999}. This results in a difference in the time needed to solve the linear problem. Fig. \ref{fig:solve_linear_system} shows the computational time of the linear solvers implemented in \textsc{scipy} and \textsc{matlab}. The default settings for the linear solvers were used for the comparison. Depending on the case, the \textsc{matlab} linear solver can take up 1.6x the time to find a solution of the problem. Nevertheless, the \textsc{matpower} implementation needs less time to find a solution than the \textsc{pypower} version, since the calculation of the Jacobian matrix is very time consuming in \textsc{pypower} v5.0.1. 

\begin{figure}[ht]
  \includegraphics[width=1.0\textwidth]{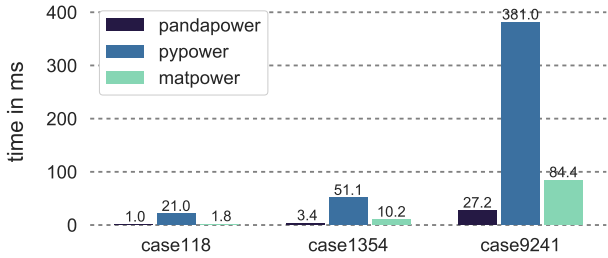}
\caption{Computational time of the Newton-Raphson method without the computational time of the linear solver}
\label{fig:newton_raphson_no_solver}       
\end{figure}

\paragraph{Linear problem}
To compare only the computational time necessary to create the linear problem, the time needed by the solver is subtracted from the total computational time. Fig. \ref{fig:newton_raphson_no_solver} shows the absolute time in ms for all cases. This figure outlines the difference in computational time achieved by the presented algorithm and the increase in speed gained through the Numba \ac{jit} compiler. Compared to the pandapower implementation, the \textsc{pypower} Newton-Raphson method needs $14x - 21x$ the time to calculate the linear problem. The \textsc{matpower} implementation needs $1.8 - 3x$ as long as the algorithm implemented in pandapower. A shorter computational time is especially relevant if multiple power flow calculations are necessary or for grids with a high quantity of buses, as in case9241. For this case, pandapower needs $27.2 ms$ to create the linear problem in comparison to $381.0 ms$ (\textsc{pypower}) and $84.4 ms$ (\textsc{matpower}).

\subsection{Concluding summary}
\begin{figure}[ht]
  \includegraphics[trim=0 0 0 0,clip, width=1.0\textwidth]{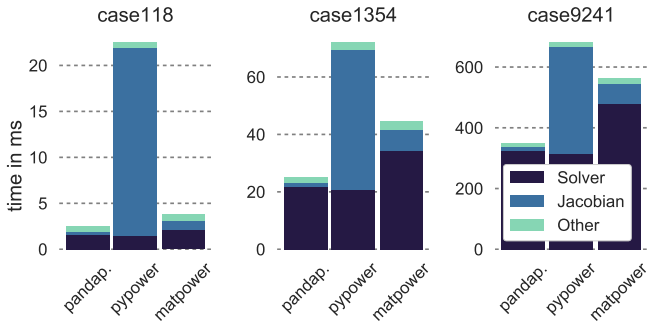}
\caption{Relative computational time of the subfunctions in the different Newton-Raphson implementations}
\label{fig:relative_timings_all}       
\end{figure}

Fig. \ref{fig:relative_timings_all} shows the relative computational time of the Newton-Raphson subfunctions implemented in pandapower, \textsc{matpower} and \textsc{pypower}. $Jacobian$ is the combined computational time for calculating the voltage derivatives and to create Jacobian matrix from these. $Solver$ is the time needed to solve the linear problem and $Other$ the computational time of all other functions. It can be seen, that 48\% of the computational time results from solving the linear problem in case118 in pandapower. Since number of buses is low, an implementation with dense matrices may decrease the calculation time. In the cases with a higher amount of buses (cases1354 and cases9241) the majority (82\% - 91\%) of calculation time is spent in the sparse solver function in pandapower. Compared to \textsc{pypower} v5.0.1 a reduction in calculation time of the Jacobian matrix of over 95\% is achieved. The speed difference in relation to \textsc{matpower} v6.02b partly results from the different linear solver. However, a reduction of calculation time to create the Jacobian matrix of at least 60\% is achieved by the presented algorithm in combination with the numba \ac{jit}-compiler. Solving the linear problem is now the main bottleneck for the presented Newton-Raphson implementation. To further reduce the time to find a solution, the Jacobian matrix ordering could be optimized for the linear solver.

\section{Conclusion}
\label{sec:conclusion}
A fast computation of power systems is crucial for modern planning and operational studies in industry as well as in research applications. A common method to find a solution to the nonlinear problem is the Newton-Raphson method, which is based on the linearization of the problem by creating the Jacobian matrix. Common implementations \cite{Lincoln.2015}\cite{Zimmerman.2011} are based on generic functions to calculate the entries of this matrix. Such implementations however cannot fully exploit the structure of the matrices which describe the power flow problem. Therefore, unnecessary iterations over the entries of these matrices are executed. To reduce the number of these iterations, a tailored algorithm based on the \ac{CRS} matrix storage format was developed and are presented in this paper. 

Comparisons of the developed method to other implementations \cite{Lincoln.2015}\cite{Zimmerman.2011} show that the total time to find a solution is less than 30 - 50 \% for the Newton-Raphson method in total. The computational effort for creating the Jacobian matrix is reduced by factors from 3x-14x, depending on the case and compared implementation. This is achieved by combining multiple matrix operations in loops and exploiting sparse techniques as well as the usage of the numba \ac{jit}-compiler. The speedup is gained without the necessity of applying additional simplifications to the power flow problem and without a loss in precision. Approaches which reduce the number of Jacobian update steps of the Newton-Raphson method \cite{Semlyen.2001} are compatible with the developed algorithm.

\section{Acknowledgement}
The research is part of the project "OpSimEval" and funded by the \ac{BMWI} (funding number 0325593B).



%
\bibliographystyle{spphys}       
\bibliography{citavi}   

\begin{acronym}
\acro{BMWI}{Federal Ministry for Economic Affairs and Energy}
\acro{CCS}{Compressed Column Storage}
\acro{CRS}{Compressed Row Storage}
\acro{nnz}{number of nonzero}
\acro{DG}{distributed generator}
\acro{DSO}{distribution system operator}
\acro{IT}{information technology}
\acro{jit}{just in time}
\acro{ICT}{information and communication technology}
\acro{KPI}{Key Performance Indicator}
\acro{OPEX}{operating expenditures}
\acro{PDP}{power distribution planning}
\acro{RES}{renewable energy resource}
\acro{SCADA}{supervisory control and data acquisition}
\end{acronym}

\end{document}